\documentclass[twocolumn,preprintnumbers,a4paper,
superscriptaddress,floatfix,twoside,prl]{revtex4}

\usepackage{bm}
\usepackage{epsfig}
\usepackage{graphics}
\usepackage{natbib}

\usepackage{fancyh}
\usepackage[l]{floatflt}
\usepackage{epsfig}
\usepackage{amssymb}
\usepackage{latexsym}
\usepackage{times}
\usepackage{slashed}
\usepackage{upgreek}
\usepackage{amsmath}
\usepackage{amsfonts}
\usepackage{amsbsy}
\usepackage{amscd}
\usepackage{bbm}

\newcommand{\Eref}[1]{Eq.~(\ref{#1})}

\newcommand{\Fref}[1]{Fig.~\ref{#1}}
\newcommand{\Tref}[1]{Table~\ref{#1}}
\newcommand{\cref}[1]{Ref.~\cite{#1}}

\newcommand{\arxiv}[1]{{\ftn\tt  arXiv:#1}}

\newcommand{\bal}{\begin{align}}
\newcommand{\eal}{\end{align}}
\newcommand{\beqs}{\begin{subequations}}
\newcommand{\eeqs}{\end{subequations}}
\newcommand{\eec}{\end{center}}
\newcommand{\bec}{\begin{center}}

\newcommand{\eem}{\end{matrix}}
\newcommand{\bem}{\begin{matrix}}
\newcommand{\eeq}{\end{equation}}
\newcommand{\beq}{\begin{equation}}
\newcommand{\ba}{\begin{array}}
\newcommand{\ea}{\end{array}}
\newcommand{\bea}{\begin{eqnarray}}
\newcommand{\eea}{\end{eqnarray}}
\newcommand{\baq}{\begin{eqnarray}}
\newcommand{\eaq}{\end{eqnarray}}

\newcommand\eqs[2]{Eqs.~(\ref{#1}) and (\ref{#2})}

\newcommand{\ftn}{\footnotesize}

\newcommand{\etal}{{\it et al.\/}}

\def\to{\rightarrow}

\def\lf{\left(}
\def\rg{\right)}
\newcommand\vev[1]{\langle {#1} \rangle}

\newcommand{\Vhi}{\ensuremath{\widehat V_{\rm CI}}}
\newcommand{\Vjhi}{\ensuremath{V_{\rm CI}}}
\newcommand{\Hhi}{\ensuremath{\widehat H_{\rm CI}}}

\newcommand{\Khi}{\ensuremath{K}}
\newcommand{\Whi}{\ensuremath{W}}
\newcommand{\Vhio}{\ensuremath{\widehat V_{\rm CI0}}}
\newcommand{\Ns}{\ensuremath{{\what N_\star}}}
\newcommand{\ck}{\ensuremath{c_{\rm K}}}

\newcommand{\mP}{\ensuremath{m_{\rm P}}}

\def\openone{\leavevmode\hbox{\small1\kern-3.8pt\normalsize1}}
\newcommand{\dV}{\ensuremath{\Delta\widehat V_{\rm CI}}}

\newcommand{\fr}{\ensuremath{f_{\cal R}}}
\newcommand{\fns}{\ensuremath{f_{n\star}}}
\newcommand{\fms}{\ensuremath{f_{m\star}}}

\newcommand{\fk}{\ensuremath{f_{\rm K}}}
\newcommand{\hk}{\ensuremath{F_{\rm K}}}
\newcommand{\hr}{\ensuremath{F_{\cal R}}}

\newcommand{\kx}{\ensuremath{k_S}}
\newcommand{\kpp}{\ensuremath{k_\Phi}}
\newcommand{\ksp}{\ensuremath{k_{S\Phi}}}
\newcommand{\ca}{\ensuremath{c_{\cal R}}}

\newcommand{\ks}{\ensuremath{k_\star}}
\newcommand{\ns}{\ensuremath{n_{\rm s}}}

\newcommand{\as}{\ensuremath{a_{\rm s}}}
\newcommand{\As}{\ensuremath{A_{\rm s}}}
\newcommand{\rw}{\ensuremath{r_{0.002}}}
\newcommand{\rs}{\ensuremath{r_{\mathcal{R}\rm K}}}
\newcommand{\rcc}{\ensuremath{\mathcal{R}}}
\newcommand{\rce}{\ensuremath{\widehat{\mathcal{R}}}}
\newcommand{\Ve}{\ensuremath{\widehat{V}}}

\newcommand{\what}{\ensuremath{\widehat}}

\def\bbet{{\bar\beta}}
\def\al{{\alpha}}

\def\n{\bar{n}}

\def\th{{\theta}}

\newcommand{\sg}{\ensuremath{\phi}}

\newcommand{\sgx}{\ensuremath{\phi_\star}}
\newcommand{\sgf}{\ensuremath{\phi_{\rm f}}}

\newcommand{\ld}{\ensuremath{\lambda}}

\newcommand{\Ld}{\ensuremath{\Lambda}}

\newcommand{\se}{\ensuremath{\widehat \phi}}
\newcommand{\sex}{\ensuremath{\widehat{\phi}_\star}}
\newcommand{\sef}{\ensuremath{\widehat{\phi}_{\rm f}}}
\newcommand{\geu}{\ensuremath{\widehat g}}
\newcommand{\eph}{\ensuremath{\widehat \epsilon}}
\newcommand{\ith}{\ensuremath{\widehat \eta}}

\def\Ka{K\"{a}hler potential}

\def\bcp{{\sc\small Bicep2}/{\it Keck Array}}
\newcommand{\plk}{{\it Planck}}

\renewcommand{\refname}{{\bf\scshape References}}

\renewenvironment{subequations}{%
\refstepcounter{equation}%
\setcounter{parentequation}{\value{equation}}%
  \setcounter{equation}{0}
  \ignorespaces
}{%
  \setcounter{equation}{\value{parentequation}}%
  \ignorespacesafterend
}

\begin{document}

%\preprint{UT-STPD-2/10}

\title{\bf\scshape Kinetically Modified Non-Minimal Chaotic Inflation}

\author{\scshape Constantinos Pallis\\ {\it Departament de F\'isica Te\`orica and IFIC,
Universitat de Val\`encia-CSIC, E-46100 Burjassot, SPAIN}
\\  {\sl e-mail address: }{\ftn\tt cpallis@ific.uv.es}}

%\date{\today}

%of the recent proposed models of inflation

\begin{abstract}

\noindent {\ftn \bf\scshape Abstract:} We consider
\emph{Supersymmetric} (SUSY) and non-SUSY models of chaotic
inflation based on the $\phi^n$ potential with $2\leq n\leq6$. We
show that the coexistence of a non-minimal coupling to gravity
$\fr=1+\ca\phi^{n/2}$ with a kinetic mixing of the form
$\fk=\ck\fr^m$ can accommodate inflationary observables favored by
the \bcp\ and \plk\ results for $0\leq m\leq4$ and
$2.5\cdot10^{-4}\leq\rs=\ca/\ck^{n/4}\leq1,$ where the upper limit
is not imposed for $n=2$. Inflation can be attained for
subplanckian inflaton values with the corresponding effective
theories retaining the perturbative unitarity up to the Planck
scale.
\\ \\ {\scriptsize {\sf PACs numbers: 98.80.Cq, 04.50.Kd, 12.60.Jv, 04.65.+e}
%Modified theories of gravity
\hfill {\sl\bfseries Published in} {\sl Phys. Rev. D} {\bf 91},
123508 (2015)

}
%\pacs{98.80.Cq, 11.30.Qc, 11.30.Er, 11.30.Pb, 12.60.Jv} it does not require fine tuned parameters

\end{abstract}\pagestyle{fancyplain}

\maketitle

\rhead[\fancyplain{}{ \bf \thepage}]{\fancyplain{}{\sl Kinetically
Modified non-MCI}} \lhead[\fancyplain{}{\sl C.
Pallis}]{\fancyplain{}{\bf \thepage}} \cfoot{}

\section{Introduction}  It is well-known \cite{old, nmi, roest} that the
presence of a non-minimal coupling function
\beq \label{fr} \fr(\sg)=1+\ca\sg^{n/2},  \eeq
between the inflaton $\sg$ and the Ricci scalar $\rcc$, considered
in conjunction with a monomial potential of the type
\beq \label{Vn}\Vjhi(\sg)=\ld^2\sg^n/2^{n/2},\eeq
provides, at the strong $\ca$ limit with $\sg<1$ -- in the reduced
Planck units with $\mP=M_{\rm P}/\sqrt{8\pi}=1$ --, an attractor
\cite{roest} towards the spectral index, $\ns$, and the
tensor-to-scalar ratio, $r$, respectively
\beq \label{nsnmi} \ns\simeq1-2/\Ns=0.965~~\mbox{and}~~r\simeq
12/\Ns^2=0.0036, \eeq
for $\Ns=55$ e-foldings with negligible $\ns$ running, $\as$.
Although perfectly consistent with the present combined \bcp\ and
\plk\ results \cite{plcp,gws},
\beq \label{data} \ns=0.968\pm0.0045~~\mbox{and}~~r=
0.048^{+0.035}_{-0.032}, \eeq
$r$ in \Eref{nsnmi} lies well below its central value in
\Eref{data} and the sensitivity of the present experiments
searching for primordial gravity waves -- for an updated survey
see \cite{cmbpol}. Nonetheless, this model -- called henceforth
non-\emph{minimal chaotic inflation} ({\sf\ftn MCI}) -- exhibits
also a weak $\ca$ regime, with $\sg>1$ and $\ca$-dependent
observables \cite{roest,oss} approaching for decreasing $\ca$'s
their values within MCI \cite{chaotic}. Focusing on this regime,
we would like to emphasize that solutions covering nicely the
1-$\sigma$ domain of the present data in \Eref{data} can be
achieved, even for $\sg<1$, by introducing a suitable
non-canonical kinetic mixing $\fk(\phi)$. For this reason we call
this type of non-MCI \emph{kinetically modified}. Although a new
parameter $\ck$, included in $\fk$, may take relatively high
values within this scheme, no problem with the perturbative
unitarity arises.
% in sharp contrast to the original scenario

\section{non-SUSY Framework}

Non-MCI is formulated in the \emph{Jordan frame} ({\sf\ftn JF})
where the action of $\phi$ is given by
\beq \label{action1} {\sf  S} = \int d^4 x \sqrt{-\mathfrak{g}}
\left(-\frac{\fr}{2}\rcc +\frac{\fk}{2}g^{\mu\nu}
\partial_\mu \sg\partial_\nu \sg-
\Vjhi(\sg)\right). \eeq
Here $\mathfrak{g}$ is the determinant of the background
Friedmann-Robertson-Walker metric, $g^{\mu\nu}$ with signature
$(+,-,-,-)$ and we allow for a kinetic mixing through the function
$\fk(\phi)$. By performing a conformal transformation \cite{nmi}
according to which we define the \emph{Einstein frame} ({\sf\ftn
EF}) metric $\geu_{\mu\nu}=\fr\,g_{\mu\nu}$ we can write ${\sf S}$
in the EF as follows
\beqs\beq \hspace*{-1.mm}{\sf  S}= \int d^4 x
\sqrt{-\what{\mathfrak{g}}}\left(-\frac12
\rce+\frac12\geu^{\mu\nu} \partial_\mu \se\partial_\nu \se
-\Vhi(\se)\right), \label{action} \eeq
where hat is used to denote quantities defined in the EF. We also
introduce the EF canonically normalized field, $\se$, and
potential, $\Vhi$, defined as follows:
\beq \label{VJe}
\frac{d\se}{d\sg}=J=\sqrt{\frac{\fk}{\fr}+{3\over2}\left({f_{\cal
R,\sg}\over \fr}\right)^2}~~~\mbox{and}~~~ \Vhi=
\frac{\Vjhi}{\fr^2}\,,\eeq\eeqs
where the symbol $,\phi$ as subscript denotes derivation
\emph{with respect to} ({\ftn\sf w.r.t}) the field $\phi$. In the
pure non-MCI \cite{old, nmi, roest} we take $\fk=1$ and so, as
shown from \Eref{VJe}, the role of $\fr$ in \Eref{fr} is twofold:
\begin{itemize}

\item[{\sf\ftn (i)}] it determines the canonical normalization of
$\se$; and \item[{\sf\ftn (ii)}] it controls the shape of $\Vhi$
affecting thereby the observational predictions. \end{itemize}

%attractive

Inspired by \cref{takahashi, lee}, where non-canonical kinetic
terms assist in obtaining inflationary solutions for $\sg<1$, we
liberate $\fr$ from its first role above implementing it by a
kinetic function of the form
\beq\label{fk} \fk(\sg)=\ck\fr^m~~\mbox{where}~~
\ck=(\ca/\rs)^{4/n},\eeq
with $\rs$ being introduced for later convenience. The form of
$\fk$ in \Eref{fk} is chosen so that the perturbative unitarity is
preserved up to Planck scale. Its most general form could be
$\fk=\ck\tilde f$ with $\tilde f$ being an arbitrary function such
that $\tilde f(\vev{\sg}=0)=1$ -- see below. However, the
variation of $\fk$ generated by $\tilde f$ can be covered by the
parametrization of \Eref{fk} selecting conveniently $m=\ln\tilde
f/\ln\fr$.

Plugging, finally, \eqs{fk}{Vn} into \Eref{VJe} we obtain
%\label{Jhi}
\beq \label{Vhio}
J^2=\frac{\ck}{\fr^{1-m}}+\frac{3n^2\ca^2\sg^{n-2}}{8\fr^2}\simeq\frac{\ck}{\fr^{1-m}}~~
\mbox{and}~~ \Vhi=\frac{\ld^2\sg^n}{2^{n/2}\fr^{2}},\eeq
assuming $\ck\gg\ca$. In contrast to \cref{lee} the presence of
both $\fk$ and $\fr$ plays a crucial role within our proposal.

%\newpage

\section{Supergravity Embeddings}

The supersymmetrization of the above models requires the use of
two gauge singlet chiral superfields, i.e., $z^\al=\Phi, S$, with
$\Phi$ ($\al=1$) and $S$ ($\al=2)$ being the inflaton and a
``stabilized'' field respectively. The EF action for $z^\al$'s
within \emph{Supergravity} ({\sf\ftn SUGRA}) \cite{linde1} can be
written as
\beqs \beq\label{Saction1}  \hspace*{-2.mm}{\sf S}=\int d^4x
\sqrt{-\what{ \mathfrak{g}}}\lf-\frac{1}{2}\rce +K_{\al\bbet}
\geu^{\mu\nu}\partial_\mu z^\al \partial_\nu z^{*\bbet}-\Ve\rg
\eeq
where summation is taken over the scalar fields $z^\al$, star
($^*$) denotes complex conjugation, $\Khi$ is the \Ka\ with
$K_{\al\bbet}=K_{,z^\al z^{*\bbet}}$ and
$K^{\al\bbet}K_{\bbet\gamma}=\delta^\al_{\gamma}$. Also $\Ve$ is
the EF F--term SUGRA potential given by
\beq \Ve=e^{\Khi}\left(K^{\al\bbet}(D_\al W) (D^*_\bbet
W^*)-3{\vert W\vert^2}\right),\label{Vsugra} \eeq \eeqs
where $D_\al W=W_{,z^\al} +K_{,z^\al}W$ with $\Whi$ being the
superpotential. Along the inflationary track determined by the
constraints
\beq \label{inftr} S=\Phi-\Phi^*=0,~\mbox{or}~~s=\bar s=\th=0\eeq
if we express $\Phi$ and $S$ according to the parametrization
\beq \Phi=\:{\phi\,e^{i
\th}}/{\sqrt{2}}\>\>\>\mbox{and}\>\>\>S=\:(s +i\bar
s)/\sqrt{2}\,,\label{cannor} \eeq
$V_{\rm CI}$ in \Eref{Vn} can be produced, in the flat limit, by
\beq \label{Wn} W=\ld S\Phi^{n/2}.\eeq
The form of $W$ can be uniquely determined if we impose two
symmetries: \begin{itemize} \item[{\sf\ftn  (i)}] an $R$ symmetry
under which $S$ and $\Phi$ have charges $1$ and $0$;
\item[{\sf\ftn (ii)}] a global $U(1)$ symmetry with assigned
charges $-1$ and $2/n$ for $S$ and $\Phi$.\end{itemize}

On the other hand, the derivation of $\Vhi$ in \Eref{Vhio} via
\Eref{Vsugra} requires a judiciously chosen $K$. Namely, along the
track in \Eref{inftr} the only surviving term in \Eref{Vsugra} is
\beq \label{1Vhio}\Vhi=\Ve(\th=s=\bar s=0)=e^{K}K^{SS^*}\,
|W_{,S}|^2\,.\eeq
The incorporation $\fr$ in \Eref{fr} and $\fk$ in \Eref{fk}
dictates the adoption of a logarithmic $K$ \cite{linde1} including
the functions
\beqs\beq \label{hr}
\hr(\Phi)=1+2^{\frac{n}{4}}\Phi^{\frac{n}{2}}\ca~~\mbox{and}~~\hk=(\Phi-\Phi^*)^2\,.
\eeq
Here \hr is an holomorphic function reducing to $\fr$, along the
path in \Eref{inftr}, and $\hk$ is a real function which assists
us to incorporate the non-canonical kinetic mixing generating by
$\fk$ in \Eref{fk}. Indeed, $\hk$ lets intact $\Vhi$, since it
vanishes along the trajectory  in \Eref{inftr}, but it contributes
to the normalization of $\Phi$ -- contrary to the naive kinetic
term $|\Phi|^2/3$ \cite{linde1} which influences both $J$ and
$\Vhi$ in \Eref{VJe}. Although $\hk$ is employed in \cref{roest}
too, its importance in implementing non-minimal kinetic terms
within non-MCI has not been emphasized so far. We also include in
$K$ the typical kinetic term for $S$, considering the
next-to-minimal term for stability reasons \cite{linde1} -- see
below --, i.e.
\beq F_S={|S|^2/3}-\kx{|S|^4/3}. \label{Fs}\eeq\eeqs
Taking for consistency all the possible terms up to fourth order,
$K$ is written as
\beqs\bea \nonumber && K=-3\ln\left(\frac{\ck}{2^m6}\lf\hr+
\hr^{*}\rg^{m}\hk \right.\\
&& \left.  +\frac12\lf\hr+\hr^*\rg-F_S+\frac\kpp6 \hk^2-\frac\ksp3
\hk{|S|^2}\right)\,.~~~~~\label{K12}\eea
Alternatively, if we do not insist on a pure logarithmic $K$, we
could also adopt the form
\beq
K=-3\ln\left(\frac12\lf\hr+\hr^*\rg-F_S\right)-\frac{\ck}{2^m}\frac{\hk}{\lf\hr+
\hr^{*}\rg^{1-m}}\,\cdot\label{K2}\eeq\eeqs
Note that for $m=0$ [$m=1$], $\hk$ and $\hr$ in $K$ given by
\Eref{K12} [\Eref{K2}] are totally decoupled, i.e. no higher order
term is needed. Our models, for $\ck\gg\ca$, are completely
natural in the 't Hooft sense because, in the limits $\ca\to0$ and
$\ld\to0$, the theory enjoys the following enhanced symmetries --
cf.~\cref{shift}:
\beq \Phi \to\ \Phi^*,\> \Phi \to\ \Phi+c \>\>\>\mbox{and}\>\>\> S
\to\ e^{i\alpha} S,\label{shift}\eeq
where $c$ is a real number. Therefore, the terms proportional to
$\ca$ can be regarded as a gravity-induced violation of the
symmetries above.

To verify the appropriateness of $K$ in Eqs.~(\ref{K12}) and
(\ref{K2}), we can first remark that, along the trough in
\Eref{inftr}, it is diagonal with non-vanishing elements
$K_{\Phi\Phi^*}=J^2$, where $J$ is given by \Eref{Vhio}, and
$K_{SS^*}=1/\fr$. Upon substitution of $K^{SS^*}=\fr$ and $\exp
K=\fr^{-3}$ into \Eref{1Vhio} we easily deduce that $\Vhi$ in
\Eref{Vhio} is recovered. If we perform the inverse of the
conformal transformation described in \eqs{action}{action1} with
frame function ${\Omega/3}=-\exp\lf-{K}/{3}\rg$ we end up with the
JF potential $\Vjhi=\Omega^2\Vhi/9$ in \Eref{Vn}. Moreover, the
conventional Einstein gravity at the SUSY vacuum,
$\vev{S}=\vev{\Phi}=0$, is recovered since $-\vev{\Omega}/3=1$.

\begin{table}[t!]
\caption{\normalfont Mass spectrum along the path in
\Eref{inftr}.}
\begin{ruledtabular}
\begin{tabular}{c|c|c}
%\begin{tabular}{c|@{\hspace{0.1cm}}c@{\hspace{0.1cm}}|@{\hspace{0.1cm}} c}\toprule
%
{\sc Fields}&{\sc Eingestates} & {\sc Mass Squared}\\ \hline
%
%\multicolumn{3}{c}{Bosons}\\ \hline
%
$1$ real scalar &$\what \th$ & $\what m^2_{\th}\simeq n_\th\Vhi/3=n_\th \Hhi^2$\\
$2$ real scalars &$\what{s},\what{\bar s}$ & $\what m^2_{s}\simeq2(6\kx\fr-1)\Hhi^2$\\
\hline\\[-0.4cm]
%
%\multicolumn{3}{c}{Fermions}\\ \hline
%
$2$ Weyl spinors&$({\what{\psi}_{S}\pm
\what{\psi}_{\Phi})/\sqrt{2}}~~$& $\what{m}^2_{\psi\pm}\simeq 3n^2
\Hhi^2/2\ck\sg^2\fr^{1+m}$
%\\\botrule
\end{tabular}
\end{ruledtabular}\label{tab1}
\end{table}
%(hatted)
%

Defining the canonically normalized fields via the relations
\beq  \label{Jg} {d\widehat \sg/
d\sg}=\sqrt{K_{\Phi\Phi^*}}=J,\>\>\> \what{\th}= J\th\sg,\eeq
and $(\what s,\what{\bar s})=\sqrt{K_{SS^*}} {(s,\bar s)}$ we can
verify that the configuration in \Eref{inftr} is stable w.r.t the
excitations of the non-inflaton fields. Taking the limit
$\ck\gg\ca$ we find the expressions of the masses squared $\what
m^2_{\chi^\al}$ (with $\chi^\al=\th$ and $s$) arranged in
\Tref{tab1}, which approach rather well the quite lengthy, exact
expressions taken into account in our numerical computation. These
expressions assist us to appreciate the role of $\kx>0$ in
retaining positive $\what m^2_{s}$. Also we confirm that $\what
m^2_{\chi^\al}\gg\Hhi^2=\Vhio/3$ for $\sgf\leq\sg\leq\sgx$ -- note
that $n_\th=4$ or $6$ for $K$ taken by \Eref{K12} or \Eref{K2},
respectively. In \Tref{tab1} we display the masses $\what
m^2_{\psi^\pm}$ of the corresponding fermions too. We define
$\what\psi_{S}=\sqrt{K_{SS^*}}\psi_{S}$ and
$\what\psi_{\Phi}=\sqrt{K_{\Phi\Phi^*}}\psi_{\Phi}$ where
$\psi_\Phi$ and $\psi_S$ are the Weyl spinors associated with $S$
and $\Phi$ respectively.

Inserting the derived mass spectrum in the well-known
Coleman-Weinberg formula, we  can find the one-loop radiative
corrections, $\dV$ to $\Vhi$. It can be verified that our results
are immune from $\dV$, provided that the renormalization group
mass scale $\Lambda$, is determined by requiring $\dV(\sgx)=0$ or
$\dV(\sgf)=0$. The possible dependence of our results on the
choice of $\Lambda$ can be totally avoided if we confine ourselves
to $\ksp\sim1$ and $\kx\sim(0.5-1.5)$ resulting to
$\Ld\simeq(4-20)\cdot10^{-5}$ -- cf. \cref{nmi, nIG}. Under these
circumstances, our results in the SUGRA set-up can be exclusively
reproduced by using $\Vhi$ in \Eref{Vhio}.

\section{Inflation Analysis}

The period of slow-roll non-MCI is determined in the EF by the
condition:
\beqs\beq{\ftn\sf
max}\{\eph(\phi),|\ith(\phi)|\}\leq1,\label{srcon}\eeq where the
slow-roll parameters $\eph$ and $\ith$ read
\beq\label{sr}\widehat\epsilon= \left({\Ve_{\rm
CI,\se}/\sqrt{2}\Ve_{\rm CI}}\right)^2
\>\>\>\mbox{and}\>\>\>\>\>\widehat\eta={\Ve_{\rm
CI,\se\se}/\Ve_{\rm CI}} \eeq\eeqs
%
%
%\beq{\ftn\sf max}\{\eph,|\ith|\}\leq1~\mbox{with}\>\>\eph=
%\frac{\Ve^2_{\rm CI,\se}}{2\Ve_{\rm CI}^2}
%\>\>\>\mbox{and}\>\>\>\ith=\frac{\Ve_{\rm CI,\se\se}}{\Ve_{\rm
%CI}},\label{srcon}\eeq
%
and can be derived employing $J$ in \Eref{VJe}, without express
explicitly $\Vhi$ in terms of $\se$. Our results are
\beq\label{sr1}\eph=\frac{n^2}{2 \sg^2\ck \fr^{1+m}};~~
\frac{\ith}{\eph}=2\lf1-\frac1n\rg-\frac{4+n(1+m)}{2n}
\ca\sg^\frac{n}{2}\,.\eeq
%
%\frac{1-n}{n}
Given that $\sg\ll1$ and so $\fr\simeq1$, \Eref{srcon} is
saturated at the maximal $\sg$ value, $\sgf$, from the following
two values
\beq \sg_{1\rm f}\simeq{n/\sqrt{2\ck}}\>\>\>\mbox{and}\>\>\>
\sg_{2\rm f}\simeq\sqrt{(n-1)n/\ck}, \label{sgf}\eeq
where $\sg_{1\rm f}$ and  $\sg_{2\rm f}$ are such that
$\eph\lf\sg_{1\rm f}\rg\simeq1$ and $\ith\lf\sg_{2\rm
f}\rg\simeq1$.

\renewcommand{\arraystretch}{1.1}
\begin{table*}[t!]
\caption{\normalfont  Inflationary predictions for $n=4$ and
$m=1,2,$ and $4$.}
\begin{ruledtabular}
\begin{tabular}{c|c|c|c}
%\begin{tabular}{c|@{\hspace{0.1cm}}c@{\hspace{0.1cm}}|@{\hspace{0.1cm}} c}\toprule
%
&{$m=1$}&$m=2$&$m=4$\\ \hline
$\ns$ &$1 - 3/2 \Ns -3/8(\Ns^3\rs)^{1/2}$&$1 - 4/3 \Ns
-1/2(3\Ns^4\rs)^{1/3}$&
$1 - 6/5 \Ns -3/5(40\Ns^6\rs)^{1/5}-3/10(50\Ns^7\rs^2)^{1/5}$\\
$r$ &$1/2 \Ns^2 \rs +2/(\Ns^3\rs)^{1/2}$&$8/3(3\Ns^4\rs)^{1/3}+4/3(9\Ns^5\rs^2)^{1/3}$&$8(4/5\Ns^6\rs)^{1/5}/5+4(16/25\Ns^7\rs^2)^{1/5}/5$\\
$\as$
&$-3/2\Ns^2-9/16(\Ns^5\rs)^{1/2}$&$-4/3\Ns^2-2/3(3\Ns^7\rs)^{1/3}$&$-6/5\Ns^2-9(4/5\Ns^{11}\rs)^{1/5}/25$
\end{tabular}
\end{ruledtabular}\label{tab2}
\end{table*}

The number of e-foldings $\Ns$ that the scale $\ks=0.05/{\rm Mpc}$
experiences during this non-MCI and the amplitude $\As$ of the
power spectrum of the curvature perturbations generated by $\sg$
can be computed using the standard formulae
\begin{equation}
\label{Nhi}  \Ns=\int_{\sef}^{\sex} d\se\frac{\Vhi}{\Ve_{\rm
CI,\se}}~~\mbox{and}~~\As^{1/2}= \frac{1}{2\sqrt{3}\, \pi} \;
\frac{\Ve_{\rm CI}^{3/2}(\sex)}{|\Ve_{\rm CI,\se}(\sex)|},\eeq
where $\sgx~[\sex]$ is the value of $\sg~[\se]$ when $\ks$ crosses
the inflationary horizon. Since $\sgx\gg\sgf$, from \Eref{Nhi} we
find
\begin{equation}
\label{Nhia} \Ns=\frac{\ck\sgx^2}{2n}
\;{}_2F_1\lf-m,4/n;1+4/n;-\ca\sgx^{n/2}\rg,
\end{equation}
%\Ns={\ck\sgx^2}
%{2n}\;{}_2F_1\lf-m,\frac4n;1+\frac4n;-\ca\sgx^{n/2}\rg. that $\Ns$
where ${}_2F_1$ is the Gauss hypergeometric function
\cite{wolfram} which reduces to unity for $m=0$ (and any $n$) or
to the factor $(\fr^{1+m}-1)/\sgx^2\ca(1+m)$ for $n=4$ (and any
$m$). Concetrating on these cases, we solve \Eref{Nhia} w.r.t
$\sgx$ with result
\begin{equation}
\label{sgx}\sgx\simeq\begin{cases}\sqrt{2n\Ns/\ck}~~\mbox{for}~~m=0,\\
\sqrt{\fms-1}/\sqrt{\rs\ck}~~\mbox{for}~~n=4,
\end{cases}
\end{equation}
where  $\fms^{1+m}=1+8(m+1)\rs\Ns$. In both cases there is a lower
bound on $\ck$, above which $\sgx<1$ and so, our proposal can be
stabilized against corrections from higher order terms. From
\Eref{Nhi} we can also derive a constraint on $\ld$ and $\ck$ i.e.
\beq \label{lan}\ld = \sqrt{3\As}\pi\cdot
\begin{cases}\lf\ck/n\Ns\rg^\frac{n}{4}\lf2n\fns/\Ns\rg^\frac12~~\mbox{for}~~m=0,\\
16\ck\rs^{3/2}/(\fms-1)^\frac32\fms^{\frac{m-1}{2}}~~\mbox{for}~~n=4
\end{cases}\eeq
where $\fns=\fr(\sgx)=1 +\rs(2n\Ns)^{n/4}$.

The inflationary observables are found from the relations
\beqs\bea \label{ns} && \ns=\: 1-6\widehat\epsilon_\star\ +\
2\widehat\eta_\star,~~r=16\widehat\epsilon_\star, \\ \label{as} &&
\as =\:2\left(4\widehat\eta_\star^2-(n_{\rm
s}-1)^2\right)/3-2\widehat\xi_\star, \eea\eeqs
where the variables with subscript $\star$ are evaluated at
$\sg=\sgx$ and $\widehat\xi={\Ve_{\rm CI,\widehat\phi} \Ve_{\rm
CI,\widehat\phi\widehat\phi\widehat\phi}/\Ve^2_{\rm CI}}$. For
$m=0$ we find
\beqs\bea\label{ns1}&& \hspace*{-2.5mm}\ns=1 - \lf4 + n
+n/\fns\rg/4\Ns,~~r=4 n/\fns\Ns,~~~~~~~~~~~~\\ &&
\hspace*{-2.5mm}\as=
\big(n^2-n(n+4)\fns-4(n+4)\fns^2\big)/16\fns^2\Ns^2\,.~~~
\label{as1}\eea\eeqs
In the limit $\rs\to0$ or $\fns\to1$ the results of the simplest
power-law MCI, \Eref{Vn}, are recovered -- cf.~\cref{chaotic}. The
formulas above are also valid for the original non-MCI
\cite{roest} with $\ck=1$ and $\rs=\ca$ lower than the one needed
to reach the attractor's values in \Eref{nsnmi}. In this limit our
results are in agreement with those displayed in \cref{oss} for
$n=4$. Furthermore, for $n=4$ (and any $m$) we obtain
\beqs\bea\label{ns2}&& \hspace*{-2.8mm}\ns=1 - 8\rs\frac{
m-1+(m+2)\fms}{(\fms-1)\fms^{1+m}},\\&& \hspace*{-2.8mm}\nonumber
r=\frac{128\rs}{(\fms-1)\fms^{1+m}},~~\as=\frac{64\rs^2(1+m)(m+2)}{(\fms-1)^2\fms^{4(1+m)}}\cdot\\
&&
\hspace*{-2.8mm}\fms^2\lf\fms^{2m}\lf\frac{1-m}{m+2}+\frac{2m-1}{m+1}\fms\rg-\fms^{2(1+m)}\rg.~~~
\label{as2}\eea\eeqs
For $n=4$ and $m=1,2$ and $4$ the outputs of
Eqs.~(\ref{ns1})-(\ref{as2}) are specified in \Tref{tab2} after
expanding the relevant formulas for $1/\Ns\ll1$.  We can clearly
infer that increasing $m$ for fixed $\rs$, both $\ns$ and $r$
increase. Note that this formulae, based on \Eref{sgx}, is valid
only for $\rs>0$ (and $m\neq0$).

From the analytic results above, see \Eref{lan} and
Eqs.~(\ref{ns1}) -- (\ref{as2}), we deduce that the free
parameters of our models, for fixed $n$ and $m$, are $\rs$ and
$\ld/\ck^{n/4}$ and not $\ck$, $\ca$ and $\ld$ as naively
expected. This fact can be understood by the following
observation: If we perform a rescaling $\sg=\tilde\sg/\sqrt{\ck}$,
\Eref{action1} preserves its form replacing $\sg$ with $\tilde\sg$
and $\fk$ with $\fr^m$ where $\fr$ and $\Vjhi$ take, respectively,
the forms
\beq\label{frVrs} \fr=1+\rs\tilde\sg^{n/2}~~\mbox{and}~~
\Vjhi=\ld^2\tilde\sg^{n}/2^{n/2}\ck^{n/2},\eeq
which, indeed, depend only on $\rs$ and $\ld^2/\ck^{n/2}$.

The conclusions above can be verified and extended to others $n$'s
and $m$'s numerically. In particular, confronting the quantities
in \Eref{Nhi} with the observational requirements \cite{plcp}
\begin{equation}
\label{Prob}
\Ns\simeq55~~\mbox{and}~~\As^{1/2}\simeq4.627\cdot10^{-5},\eeq
we can restrict $\ld/\ck^{n/4}$ and $\sgx$ and compute the model
predictions via \eqs{ns}{as}, for any selected $m, n$ and $\rs$.
The outputs, encoded as lines in the $\ns-\rw$ plane, are compared
against the observational data \cite{plcp,gws} in \Fref{fig1} for
$m=0,1,2,$ and $4$ and $n=2$ (dashed lines), $n=4$ (solid lines),
and $n=6$ (dot-dashed lines). The variation of $\rs$ is shown
along each line. To obtain an accurate comparison, we compute
$\rw=16\eph(\sg_{0.002})$ where $\sg_{0.002}$ is the value of
$\sg$ when the scale $k=0.002/{\rm Mpc}$, which undergoes $\what
N_{0.002}=(\Ns+3.22)$ e-foldings during non-MCI, crosses the
horizon of non-MCI.

\begin{figure*}[!t]
\includegraphics[width=60mm,angle=-90]{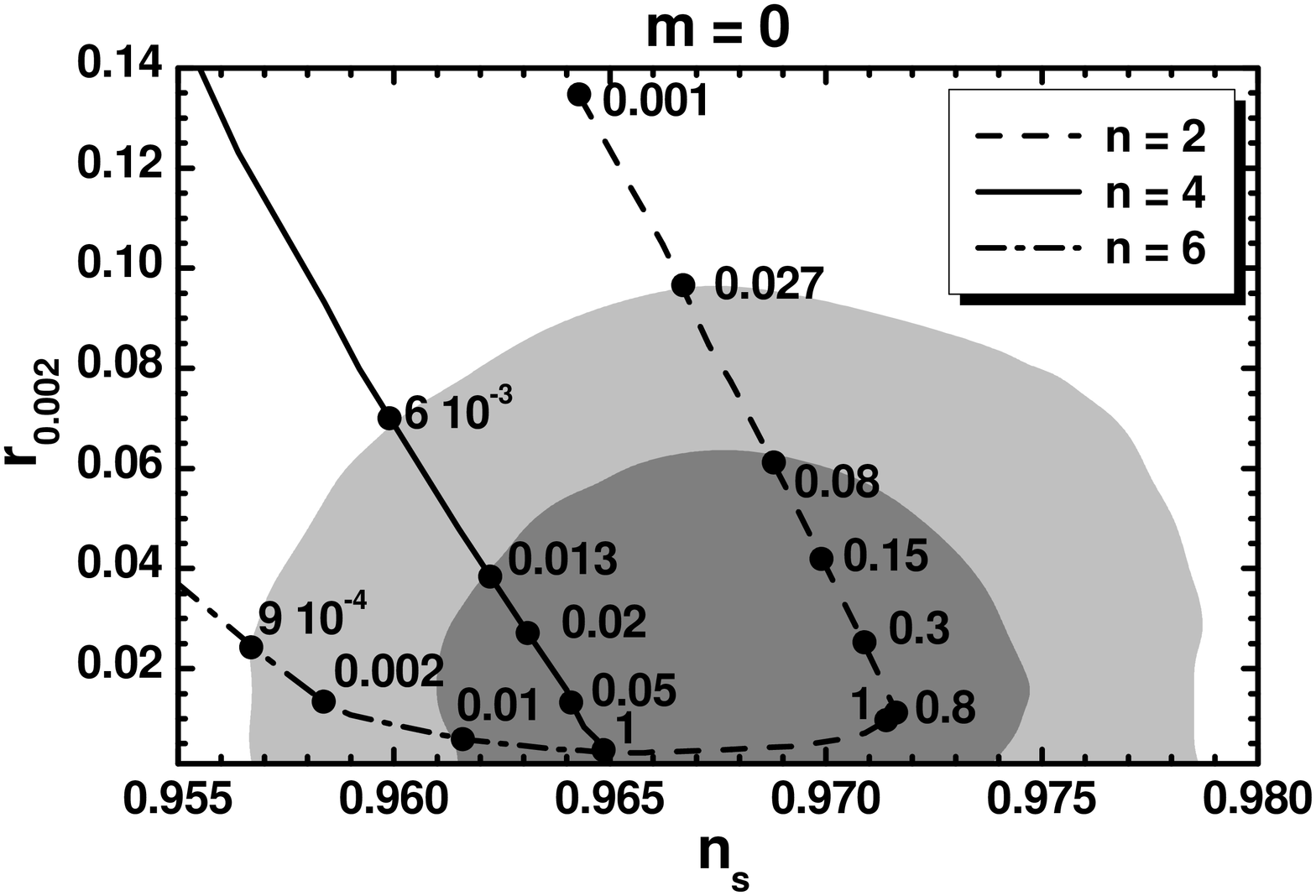}
\includegraphics[width=60mm,angle=-90]{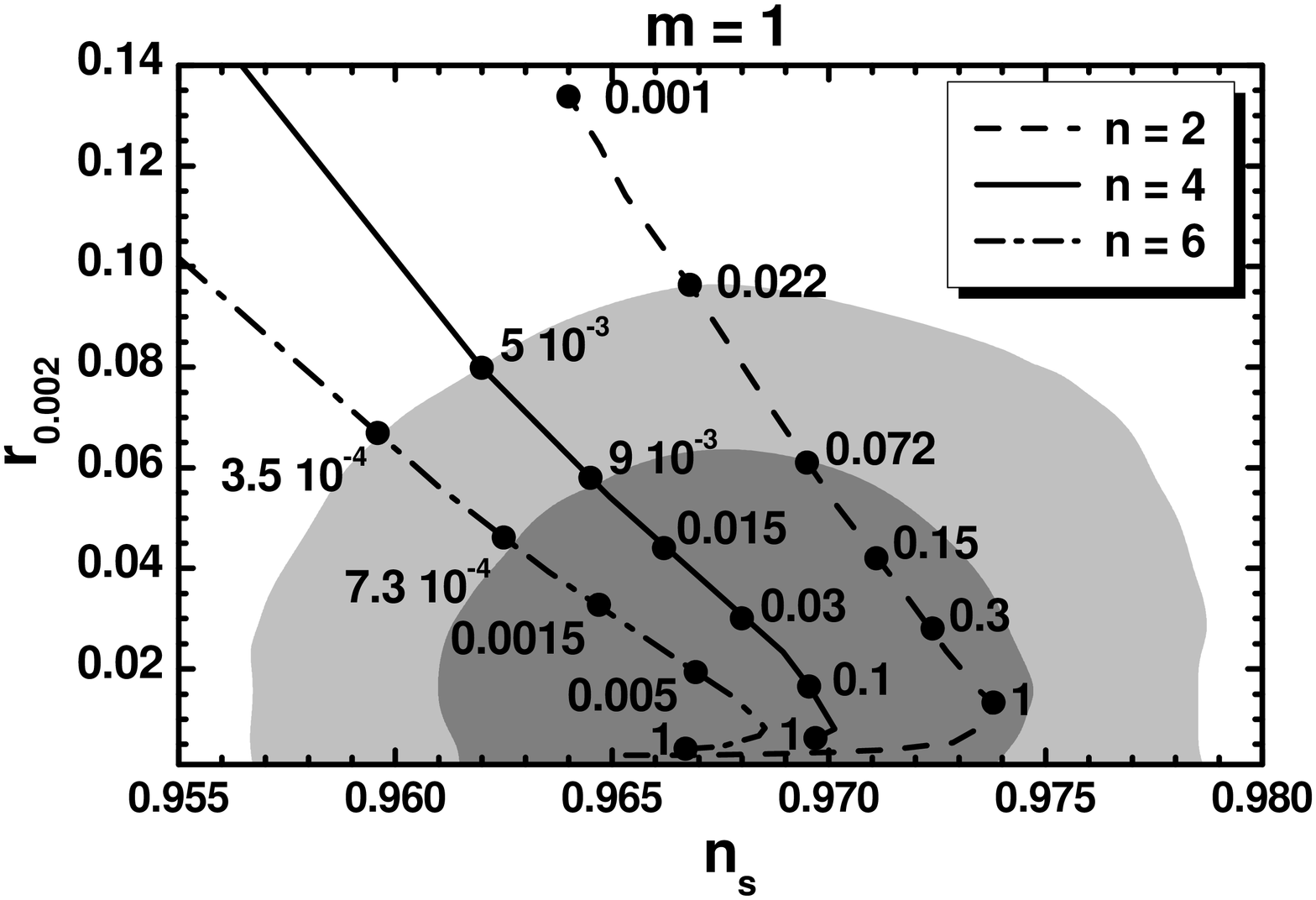}\\
\includegraphics[width=60mm,angle=-90]{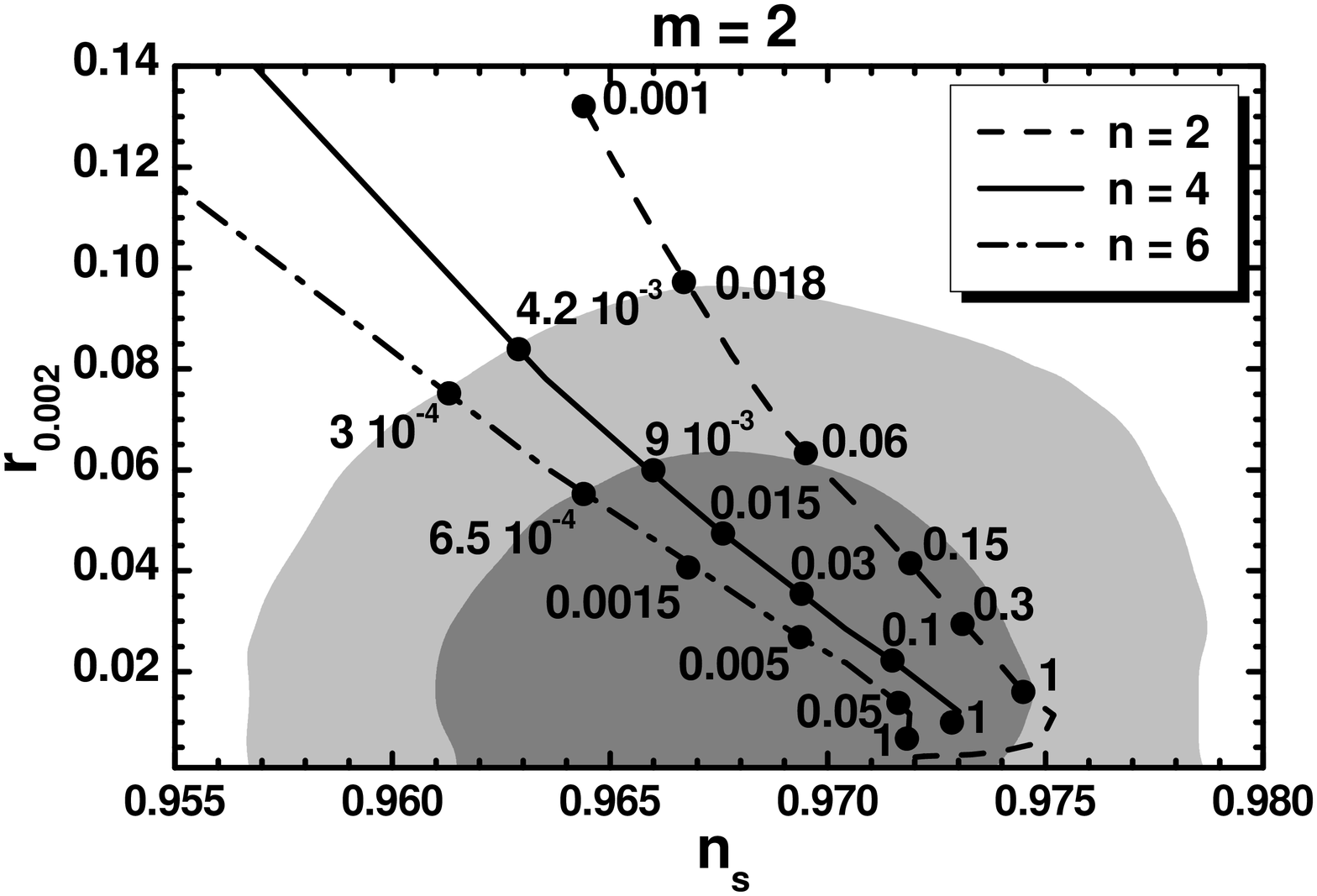}
\includegraphics[width=60mm,angle=-90]{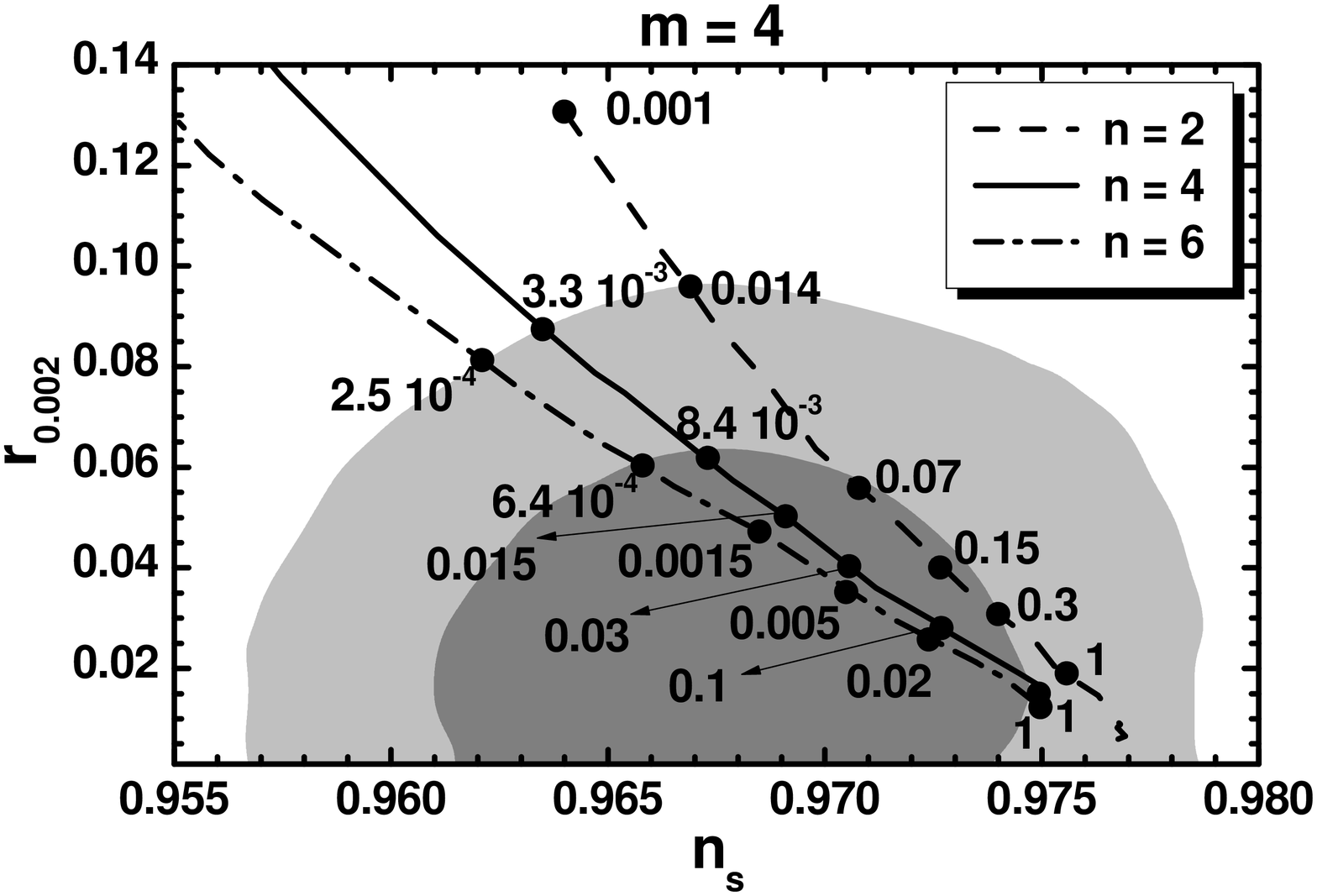}
\caption{\sl Allowed curves in the $\ns-\rw$ plane for $m=0,1,2$
and $4$, $n=2$ (dashed lines), $n=4$ (solid lines), $n=6$
(dot-dashed lines) and various $\rs$'s indicated on the curves.
The marginalized joint $68\%$ [$95\%$] regions from \plk, \bcp\
and BAO data are depicted by the dark [light] shaded
contours.}\label{fig1}
\end{figure*}

From the plots in \Fref{fig1} we observe that, for low enough
$\rs$'s -- i.e. $\rs=10^{-7}, 10^{-4},$ and $0.001$ for $n=6, 4,$
and $2$ --, the various lines converge to the $(\ns,\rw)$'s
obtained within MCI. At the other end, the lines for $n=4$ and $6$
terminate for $\rs=1$, beyond which the theory ceases to be
unitarity safe -- see below -- whereas the $n=2$ line approaches
an attractor value for any $m$. For $m=0$ we reveal the results of
\cref{roest}, i.e. the displayed lines are almost parallel for
$\rw\geq0.02$ and converge at the values in \Eref{nsnmi} -- for
$n=4$ and $6$ this is reached even for $\rs=1$. For $m>0$ the
curves move to the right and span more densely the 1-$\sigma$
ranges in \Eref{data} for quite natural $\rs$'s -- e.g.
$0.005\lesssim\rs\lesssim0.1$ for $m=1$ and $n=4$. It is worth
mentioning that the requirement $\rs\leq1$ provides a lower bound
on $\rw$, which ranges from $0.0032$ (for $m=0$ and $n=6$) to
$0.015$ (for $m=4$ and $n=4$). Note, finally, that our estimations
in Eqs.~(\ref{ns1})--(\ref{as1}) are in agreement with the
numerical results for $n=2$ and $\rs\lesssim1$, $n=6~[4]$ and
$\rs\lesssim0.002~[0.05]$. For $m>0$ (and $n=4$) our findings in
Eqs.~(\ref{ns2})--(\ref{as2}) (and \Tref{tab2}) approximate fairly
the numerical outputs for $0.003\lesssim\rs\leq1$.

\section{Effective Cut-Off Scale}\label{uv}

The selected $\fk$ in \Eref{fk} not only reconciles non-MCI with
the 1-$\sigma$ ranges in \Eref{data} but also assures that the
corresponding effective theories respect perturbative unitarity up
to $\mP=1$ although $\ck$ may take relatively large values for
$\sg<1$ -- e.g. for $n=4, m=1$ and $\rs=0.03$ we obtain
$140\lesssim\ck\lesssim1.4\cdot10^6$ for
$3.3\cdot10^{-4}\lesssim\ld\lesssim3.5$. This achievement stems
from the fact that $\se=\vev{J}\sg$ does not coincide -- contrary
to the pure non-MCI \cite{cutoff, riotto} for $n>2$ -- with $\sg$
at the vacuum of the theory, given that $\vev{J}=\sqrt{\ck}$ or
$\vev{J}=\sqrt{\ck+3\ca^2/2}$ for $\vev{\sg}=0$ and $n>2$ or $n=2$
-- see \Eref{Vhio}. It is notable that this by-product of our
proposal for $n>2$ arises without invoking large $\vev{\sg}$'s as
in \cref{nIG,lee,gian}.

To clarify further this point we analyze the small-field behavior
of our models in the EF. We focus on the second term in the
right-hand side of \Eref{action} or (\ref{Saction1}) for
$\mu=\nu=0$ and we expand it about $\vev{\phi}=0$ in terms of
$\se$ -- see \Eref{VJe}. Our result for $m=0$ and $n=2,4,$ and $6$
can be written as
\beqs\beq\nonumber J^2
\dot\phi^2=\lf1-\rs\what{\sg}^\frac{n}{2}+\frac{3n^2}{8}\rs^2\what{\sg}^{n-2}+
\rs^2\what{\sg}^{n}\cdots\rg\dot\se^2.\eeq
Similar expressions can be obtained for the other $m$'s too.
Expanding similarly $\Vhi$, see \Eref{Vhio}, in terms of $\se$ we
have
\beq\nonumber
\Vhi=\frac{\ld^2\what{\sg}^n}{2\ck^{n/2}}\lf1-2\rs\what{\sg}^{\frac{n}{2}}+3\rs^2\what{\sg}^n-
4\rs^3\what{\sg}^{\frac{3n}{2}}+\cdots\rg, \eeq\eeqs
independently of $m$. From the expressions above we conclude that
our models do not face any problem with the perturbative unitarity
for $\rs\leq1$. For $n=2$ this statement is also valid even for
$\rs>1$ as shown in \cref{nmi,riotto}. In the latter case, though,
the naturalness argument mentioned below \Eref{K2} is invalidated.

% $\Qef=\rs^{-2/n}\mP$ and therefore .

\section{Conclusions} Prompted by the recent
joint analysis of \bcp\ and \plk\ which, although does not exclude
inflationary models with negligible $r$'s, seems to favor those
with $r$'s of order $0.01$ we proposed a variant of non-MCI which
can safely accommodate $r$'s of this level. The main novelty of
our proposal is the consideration of the non-canonical kinetic
mixing in \Eref{fk} -- involving the parameters $m$ and $\ck$ --
apart from the non-minimal coupling to gravity in \Eref{fr} which
is associated with the potential in \Eref{Vn}. This setting can be
elegantly implemented in SUGRA too, employing the super- and \Ka s
given in \eqs{Wn}{K12} or (\ref{K2}). Prominent in this
realization is the role of a shift-symmetric quadratic function
$\hk$ in \Eref{hr} which remains invisible in the SUGRA scalar
potential while dominates the canonical normalization of the
inflaton. Using $m\geq0$ and confining $\rs$ to the range
$(2.5\cdot10^{-4}-1)$, where the upper bound does not apply to the
$n=2$ case, we achieved observational predictions which may be
tested in the near future and converge towards the ``sweet'' spot
of the present data -- its compatibility with the $m=1$ case,
especially for $n=4$ and $6$, is really impressive -- see
\Fref{fig1}. These solutions can be attained even with
subplanckian values of the inflaton requiring large $\ck$'s and
without causing any problem with the perturbative unitarity. It is
gratifying, finally, that a sizable fraction of the allowed
parameter space of our models (with $n=4$) can be studied
analytically and rather accurately.

%the corresponding effective theory respects  up to $\mP$.

%\vspace*{-.5cm}

\paragraph*{\small\bfseries\scshape Acknowledgments} {\small The author acknowledges useful
discussions with G.~Lazarides, A.~Racioppi, and G.~Trevisan. This
research was supported from the MEC and FEDER (EC) grants
FPA2011-23596 and the Generalitat Valenciana under grant
PROMETEOII/2013/017.}

%\onecolumngrid \bec\rule{0.5\textwidth}{1.pt}\eec\vspace*{-1.2cm}

\def\ijmp#1#2#3{{\sl Int. Jour. Mod. Phys.}
{\bf #1},~#3~(#2)}
\def\plb#1#2#3{{\sl Phys. Lett. B }{\bf #1}, #3 (#2)}
\def\prl#1#2#3{{\sl Phys. Rev. Lett.}
{\bf #1},~#3~(#2)}
\def\rmp#1#2#3{{Rev. Mod. Phys.}
{\bf #1},~#3~(#2)}
\def\prep#1#2#3{{\sl Phys. Rep. }{\bf #1}, #3 (#2)}
\def\prd#1#2#3{{\sl Phys. Rev. D }{\bf #1}, #3 (#2)}
\def\npb#1#2#3{{\sl Nucl. Phys. }{\bf B#1}, #3 (#2)}
\def\npps#1#2#3{{Nucl. Phys. B (Proc. Sup.)}
{\bf #1},~#3~(#2)}
\def\mpl#1#2#3{{Mod. Phys. Lett.}
{\bf #1},~#3~(#2)}
\def\jetp#1#2#3{{JETP Lett. }{\bf #1}, #3 (#2)}
\def\app#1#2#3{{Acta Phys. Polon.}
{\bf #1},~#3~(#2)}
\def\ptp#1#2#3{{Prog. Theor. Phys.}
{\bf #1},~#3~(#2)}
\def\n#1#2#3{{Nature }{\bf #1},~#3~(#2)}
\def\apj#1#2#3{{Astrophys. J.}
{\bf #1},~#3~(#2)}
\def\mnras#1#2#3{{MNRAS }{\bf #1},~#3~(#2)}
\def\grg#1#2#3{{Gen. Rel. Grav.}
{\bf #1},~#3~(#2)}
\def\s#1#2#3{{Science }{\bf #1},~#3~(#2)}
\def\ibid#1#2#3{{\it ibid. }{\bf #1},~#3~(#2)}
\def\cpc#1#2#3{{Comput. Phys. Commun.}
{\bf #1},~#3~(#2)}
\def\astp#1#2#3{{Astropart. Phys.}
{\bf #1},~#3~(#2)}
\def\epjc#1#2#3{{Eur. Phys. J. C}
{\bf #1},~#3~(#2)}
\def\jhep#1#2#3{{\sl J. High Energy Phys.}
{\bf #1}, #3 (#2)}
\newcommand\jcap[3]{{\sl J.\ Cosmol.\ Astropart.\ Phys.\ }{\bf #1}, #3 (#2)}
\newcommand\njp[3]{{\sl New.\ J.\ Phys.\ }{\bf #1}, #3 (#2)}


\begin{thebibliography}{99}
 \section*{\refname}  %\ignorespaces

\bibitem{old}  D. S. Salopek, J. R. Bond and J.M.
Bardeen, {\sl Phys. Rev. D }{\bf 40}, 1753 (1989); F.L.~Bezrukov
and M.~Shaposhnikov, \plb{659}{2008}{703}  [\arxiv{0710.3755}].

\bibitem{nmi} C. Pallis, \plb{692}{2010}{287} [\arxiv{1002.4765}];
C. Pallis and Q. Shafi, \prd{86}{2012}{023523} [\arxiv{
1204.0252}]; C. Pallis and Q. Shafi, \jcap{03}{2015}{023}
[\arxiv{1412.3757}].
%%CITATION = ARXIV:1412.3757;%%
%%CITATION = ARXIV:1204.0252;%%
%%CITATION = ARXIV:1002.4765;%%


\bibitem{roest} R. Kallosh, A. Linde, and D. Roest,
{\sl Phys. Rev. Lett.} {\bf 112}, 011 303 (2014)
[\arxiv{1310.3950}].

\bibitem{plcp} \plk\ Collaboration, \arxiv{1502.02114}.
%%``Planck 2015 results. XIII. Cosmological parameters,'
  %%CITATION = ARXIV:1502.02114;%%


\bibitem{gws} P.A.R.~Ade {\it et al.}  [\bcp\ and \plk\ Collaborations],
\prl{114}{2015}{101301} [\arxiv{1502. 00612}].
  %%CITATION = ARXIV:1502.00612;%%

%\bibitem{plnext} J. Tauber \etal\ [\plk\ Collaboration],
%\astroph{0604069}.

\bibitem{cmbpol} P.~Creminelli \etal, \arxiv{1502.01983}.

\bibitem{oss} N.~Okada, M.U.~Rehman, and Q.~Shafi, \prd{82}{2010}{043502} [\arxiv{1005.5161}];
N.~Okada, V.N. \c{S}eno\u{g}uz, and Q.~Shafi, \arxiv{1403.6403}.

\bibitem{chaotic} A.D. Linde, \plb{129}{1983}{177}.

\bibitem{takahashi} F.~Takahashi, {\sl Phys. Lett. B}
{\bf 693}, 140 (2010) [\arxiv{1006. 2801}]; K.~Nakayama and
F.~Takahashi, \jcap{11}{2010}{009} [\arxiv{1008.2956}].


\bibitem{lee} H.M.~Lee, {\sl Eur.\ Phys.\ J.\ C }{\bf 74}, 3022 (2014)
[\arxiv{1403. 5602}].
  %%CITATION = ARXIV:1403.5602;%%
  %``Chaotic inflation and unitarity problem,''




\bibitem{linde1} M.B.~Einhorn and D.R.T.~Jones,
\jhep{03}{2010}{026} [\arxiv{0912.2718}]; H.M.~Lee,
\jcap{08}{2010}{003} [\arxiv{1005.2735}]; S.~Ferrara \etal,
\prd{83}{2011}{025008} [\arxiv{1008.2942}]; C.~Pallis and
N.~Toumbas, \jcap{02}{2011}{019} [\arxiv{1101.0325}].
%%CITATION = ARXIV:1101.0325;%%
%%CITATION = ARXIV:1108.1771;%%

\bibitem{shift}  R.~Kallosh, A.~Linde, and T.~Rube,
\prd{83}{2011}{043507} [\arxiv{1011.5945}].

\bibitem{nIG}  C.~Pallis, \jcap{04}{2014}{024}
[\arxiv{ 1312.3623}]; C.~Pallis, \jcap{08}{2014}{057}
[\arxiv{1403.5486}]; C.~Pallis, \jcap{10}{2014}{058}
[\arxiv{1407.8522}].
  %%CITATION = ARXIV:1312.3623;%%
  %%CITATION = ARXIV:1403.5486;%%
  %%CITATION = ARXIV:1407.8522;%%

\bibitem{wolfram} {\ftn\sf http://functions.wolfram.com}.

\bibitem{cutoff} J.L.F.~Barbon and J.R.~Espinosa,
\prd{79}{2009}{081302} [\arxiv{0903.0355}]; C.P.~Burgess,
H.M.~Lee, and M.~Trott, \jhep{07}{2010}{007} [\arxiv{1002. 2730}].

\bibitem{riotto} A.~Kehagias, A.M.~Dizgah, and A.~Riotto, \prd{89}{2014}{043527}
[\arxiv{1312.1155}].


\bibitem{gian} G.F. Giudice and H.M. Lee, \plb{733}{2014}{58}
[\arxiv{1402.2129}].



\end{thebibliography}
\end{document}